# High-precision 2D surface phosphor thermometry at kHz-rates during flame-wall interaction in narrow passages

Anthony O. Ojo[a,*], David Escofet-Martin[a,b], Brian Peterson[a]

[a] *Institute for Multiscale Thermofluids, School of Engineering, The University of Edinburgh, EH9 3FD, United Kingdom.*
[b] *CMT - Motores Térmicos, Universitat Politècnica de València, 46022, Valencia, Spain.*

______________________________________________________________________

**Abstract**

This work demonstrates high-speed 2D wall temperature measurements occurring during flame-wall interaction (FWI) within a narrow channel. Such measurements are essential to understand transient wall heat transfer and flame behavior occurring within micro-combustors and designated engine crevices. Wall temperature is measured using the phosphor Bismuth-doped Scandium vanadate ($ScVO_4:Bi^{3+}$). $ScVO_4:Bi^{3+}$ exhibits a short phosphorescence lifetime (~ 2 µs at room temperature), enabling kHz measurement rates. $ScVO_4:Bi^{3+}$ also exhibits a high temperature sensitivity, which yields single-shot precision < 0.5 K within the temperature range of 295 – 335 K. A frequency-doubled Ti:Sapphire laser emitting light at 400 nm is used to excite $ScVO_4:Bi^{3+}$, and wall temperature is measured within a ~ 22 × 22 mm$^2$ region with 380 µm spatial resolution. Phosphor thermometry and CH* imaging are combined at 1 kHz to measure the spatiotemporal dynamics of the flame and wall temperature ($T_{wall}$) within a 2 mm crevice passage in a fixed volume chamber designed for heat transfer studies. Measurements describe $T_{wall}$ signatures associated with transient FWI events, including unique $T_{wall}$ features associated with wrinkled flame fronts. For our operating conditions, $T_{wall}$ associated with flame cusps consistently exhibit temperatures 5 – 20 K lower than flame crest regions. The most extreme difference is seen for large cusp formation, where local wall cooling is noticeable at the trough of the flame cusp. This cooling feature may be caused by intrinsic flame/flow instabilities, which locally and temporally cool the wall.

______________________________________________________________________

*Email address:* anthony.ojo@ed.ac.uk (A. O. Ojo)     *Corresponding author.

## 1. Introduction

Understanding flame-wall interaction (FWI) and wall heat losses in narrow channels is an important topic for technically relevant combustion systems such as micro-combustors, flame arrestors, and internal combustion (IC) engines. Narrow channels are characterized by large surface-area-to-volume ratios, which yield significant gaseous heat loss and flame quenching. These processes define emissions and performance in micro-combustors and IC engines [1–3], as well as define safe operation of flame arrestors. Developing cleaner and safe combustion systems requires the ability to understand near-wall heat transfer and FWI processes in narrow channels.

Detailed measurements of heat loss and FWI in narrow channels are less common, primarily due to limited optical access in these passages. In early research, ion probes were means to understand flame quenching in narrow passages [4]. Since then, more sophisticated measurements, including flame imaging [5,6] and high-speed filtered Rayleigh [7] have provided valuable insight into FWI in narrow channels. Numerical simulations have also provided valuable insight of flame behavior in narrow channels [e.g. [5,8]. However, simulations require detailed understanding of wall boundary conditions with respect to the flame in order to resolve the flame behavior accurately. Experimental measurements of wall temperature ($T_{wall}$) are needed to understand FWI and support further development of wall modeling.

Phosphor thermometry provides a reliable means to measure surface temperature in several combustion applications [9]. The technique relies on applying a thin coating of thermographic phosphor (TGP) particles on a surface and exciting the coating with incident light at an appropriate wavelength. Surface temperature can be derived from the phosphorescence properties of the TGP using either the spectral intensity ratio or the lifetime approach. Single-point (0-D) TGP measurements are most common for surface temperature [9–12]. Recently, 0-D phosphor thermometry was used to study FWI in a narrow passage emulating a piston crevice [12]. $T_{wall}$ and FWI quantities (wall heat flux and quenching distance) were derived and referenced to the flame behavior monitored by high-speed flame imaging. Quantities were highly dependent on the flame behavior, which varied spatially and temporally. This revealed the need for high-speed, 2D thermometry to understand transient heat transfer and flame behavior.

For 2D phosphor thermometry, it is becoming attractive to employ a single camera and use a lifetime based intensity-ratio approach which relies on the ratio of intensities imaged at two intervals during the TGP's phosphorescence [13–15]. This approach reduces data processing time compared to the 2D lifetime approach, where multiple images are required to resolve the phosphorescence decay time [16,17]. Most studies implementing the lifetime-based intensity-ratio approach have utilized low recording rates (typically < 20 Hz). Recently, 2D surface thermometry measurements have been reported at 1 kHz using a novel intensity ratio [18], but has not been used to study heat transfer for FWI processes.

2D measurements at kHz rates are required to resolve the transient flame-wall dynamics occurring at short timescales within narrow passages. We present a lifetime-based intensity-ratio strategy to obtain temporally and spatially resolved 2D $T_{wall}$ measurements at 1 kHz. Phosphor thermometry is performed simultaneously with flame front (CH*) imaging in a narrow crevice passage of an optically accessible fixed-volume chamber (FVC). For thermometry, the TGP Bismuth-doped Scandium vanadate ($ScVO_4:Bi^{3+}$) is used, which has a phosphorescence lifetime < 10 μs at room temperature and offers high temperature sensitivity and precision (< 0.5 K) [12,14]. Measurements describe $T_{wall}$ signatures associated with transient FWI events, including unique $T_{wall}$ features associated with wrinkled flame fronts and local wall cooling, which the latter is likely induced by hydrodynamic instabilities.

## 2. Experimental setup

### 2.1 Fixed volume chamber (FVC)

Experiments are performed within an optically accessible FVC, shown in Fig. 1. The FVC features a test section (150 cm$^3$) and a back-pressure section (6 cm$^3$). These sections are separated by a 6 mm thick orifice plate with 81 equidistant holes each of 0.5 mm diameter. The test section emulates a simplified IC engine geometry at top-dead center. This includes a crevice region, emulating a piston crevice where heat transfer and flame quenching are most severe [5,19]. Fused silica (FS) windows provide optical access to the test section and crevice passage.

The FVC operation has been demonstrated in [11,12,20] and is briefly described below. The FVC is initially evacuated to 20 mbar. A homogeneous CH$_4$-air mixture with an equivalence ratio, $\phi$ = 1.0 is then introduced into the FVC until an initial pressure ($P_i$) is reached. The mixture is ignited via a spark plug. Heat release initiates an exponential pressure rise as the flame propagates towards the opposite end of the FVC before entering the crevice region. At a preselected pressure, a dump-valve is actuated to evacuate the FVC. The exiting exhaust flow is choked via an orifice plate, providing an exponential pressure decay. In this work, $P_i$ of 1 bar and 2 bar were employed for measurements. At these $P_i$, the preselected pressure for dump valve actuation was 2.1 bar and 4.1 bar, respectively.

This work focuses on $T_{wall}$ measurements and flame front imaging within the crevice passage of the FVC. The crevice is 70 mm deep ($\Delta y$), and 158 mm wide ($\Delta z$). The crevice region is characterized by two walls: the front FS window surface and a metal wall. The distance between both walls ($\Delta x$) is the crevice spacing (CS). The front FS window placement can be adjusted to select the crevice spacing (0.5-5 mm). Here, a fixed CS of 2mm was used for experiments. The front window ($\Delta y \times \Delta z$; 60 ×110 mm) provides optical access for the majority of the crevice region. Metal components of the FVC are manufactured from 304 stainless steel. The FVC walls are not externally cooled or heated in this work.

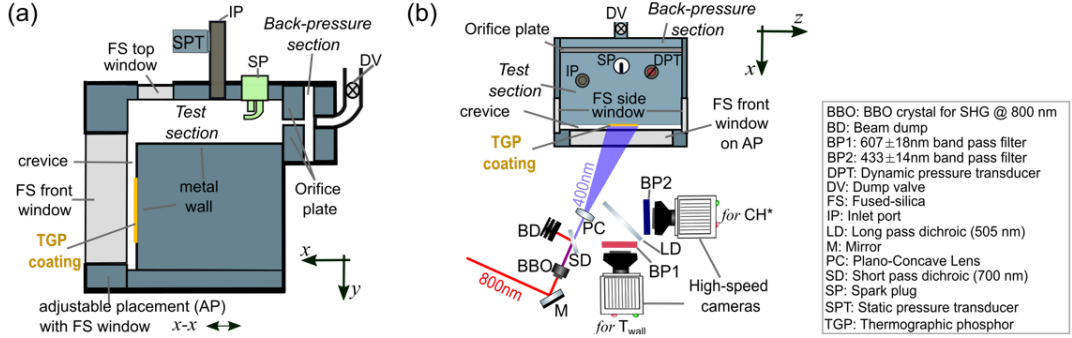

Fig. 1. (a) Schematic of FVC, (b) schematic of the experimental setup.

## 2.2 Phosphor thermometry

The TGP, bismuth-doped (1 mol %) scandium vanadate ($ScVO_4:Bi^{3+}$), which was synthesized in Ref [14], is used for surface thermometry in this work. This TGP was shown to exhibit negligible dependence on excitation laser fluence [12,14], which is optimal for optically attenuating environments [12].

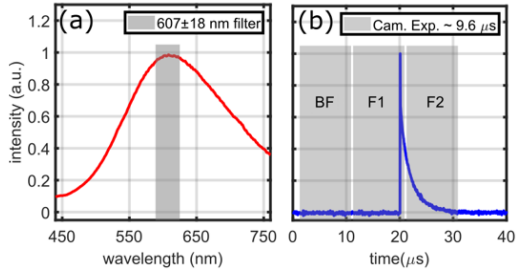

Fig. 2. $ScVO_4:Bi^{3+}$ phosphorescence (a) spectrum from [14], (b) decay signal with camera frames used for thermometry.

Figure 2a shows the room temperature emission spectrum of $ScVO_4:Bi^{3+}$. As shown, $ScVO_4:Bi^{3+}$ exhibits a broadband emission spectrum and its emission is credited to the $^3P_1 \rightarrow {}^1S_0$ and $^3P_0 \rightarrow {}^1S_0$ transitions of $Bi^{3+}$ [21,22]. The phosphorescence signals are detected within the 607±18 nm wavelength range, specified by a spectral filter. Figure 2b shows an example of the phosphorescence decay signal of $ScVO_4:Bi^{3+}$ at room temperature. It is shown that $ScVO_4:Bi^{3+}$ features a fast phosphorescence decay with a lifetime of 2 μs. The short lifetime of $ScVO_4:Bi^{3+}$ makes it possible for high-speed measurements, on the order of several 10's of kHz.

The TGP was mixed with a temperature resistant binder (HPC binder, ZYP coatings) and was applied using an airbrush to produce a ~ 40 mm × 30 mm ($\Delta y \times \Delta z$) coating on the metal wall in the crevice region (Fig. 1). The coating thickness ($\Delta x$) was measured at different spatial locations using a coating thickness (CT) gauge. The coating was ~ 7 ± 2 μm thick. The TGP coating was excited by a 400 nm laser beam generated from a Ti:Sapphire laser (800nm, 35 fs pulse width) equipped with BBO crystal for second harmonic generation. The 400 nm laser beam was expanded and its diameter incident on the TGP coating was 35 mm. The laser operated at 1 kHz with laser energy of 0.15 mJ/pulse at 400 nm.

A high-speed camera (VEO 710L, Phantom) was used to image phosphorescence from the TGP coating. Phosphorescence signals were imaged through a longpass dichroic mirror, which transmitted wavelengths > 505 nm and reflected wavelengths < 505 nm. The reflected light was used for CH* detection. For thermometry, a 50 mm Nikon camera lens (f/1.2) fitted with a 607±18 nm bandpass filter was mounted on the high-speed camera. The imaging setup provided a ~ 190 μm/pixel resolution, giving a final spatial resolution of 380 μm/pixel after 2 × 2 pixel binning. The region of interest on the TGP coating was ~ 22 × 22 mm, which was within the central part of the 35 mm diameter incident laser beam.

Operation and data acquisition by the high-speed camera was performed using the software DaVis (LaVision). Data acquisition is synchronized with the ignition timing for combustion of the $CH_4$-air mixture. The camera operated at a frame rate of 100 kHz, with an interframe time of ~ 400 ns, such that the camera exposure was ~9.6 μs. For each phosphorescence signal that follows the 1 kHz laser excitation, three frames were acquired. Figure 2b shows these frames for a phosphorescence decay signal. The first frame (BF) captures a background image. The second frame (F1) images the first ~0.85 μs of the phosphorescence decay, and the third frame (F2) images the remaining phosphorescence signal within the 9.6 μs exposure. For each decay signal, the lifetime-based intensity ratio (IR) field is evaluated on a pixel-by-pixel basis as shown by Equation (1).

$$IR = \frac{F1 - BF}{F2 - BF} \quad (1)$$

A flat-field correction was then performed to correct for the spatial variation of the laser intensity. To do this, a time-averaged IR field recorded at room temperature ($\overline{IR}_{296K}$) was obtained. Individual IR fields evaluated from images taken during temperature measurements are then normalized by $\overline{IR}_{296K}$ (pixel-by-pixel) according to Equation (2). This yields a resultant ratio field/image, R evaluated for each laser shot/phosphorescence signal.

$$R = \frac{IR}{\overline{IR}_{296K}} \quad (2)$$

It should be mentioned that the background image can be taken at the start of experiments, independently from F1 and F2, as long as it is taken in the absence of the phosphorescence signal. That is, IR for each phosphorescence signal can be evaluated using a single BF, for which R can be evaluated subsequently from Equation (2). This strategy was equally explored in this work and provided similar results to the three-frame approach described in Fig. 2b.

Calibration of R as a function of temperature was conducted outside the FVC using the same optical setup and imaging procedure used for measurements in the FVC. The TGP was applied on a surface of a small flat aluminium bar (30 × 30 × 10 mm) equipped with a thermocouple on its surface. The bar was heated in an oven to ~373 K and then removed to cool on a platform having fiberglass insulation. While monitoring the bar temperature with a thermocouple, images were taken at selected temperatures < 345 K as the bar cooled to room temperature.

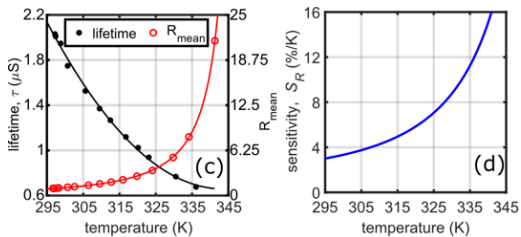

Fig. 3. (a) Phosphorescence lifetime and intensity ratio versus temperature. (b) resultant temperature sensitivity.

Figure 3a shows the phosphorescence lifetime, τ, obtained during a temperature calibration routine using a Photomultiplier Tube (PMT) to detect phosphorescence decays from a ~3 mm diameter laser-excited area on the TGP coated aluminium bar. In Fig. 3a, we present the 0-D phosphorescence lifetime data to show that τ decreases with temperature and to describe how changes in τ affect the 2D lifetime-based intensity ratio strategy adopted in this work. Figure 3a shows that τ strongly decreases in the 295 - 335 K temperature range. Due to thermal quenching, τ decreases from ~ 2 µs at 294 K to ~ 0.64 µs at 335 K.

Figure 3a also shows the temperature dependence of the spatially averaged R data ($R_{mean}$), which constitutes our calibration curve. The calibration curve is used to convert the evaluated R images to temperature images on a pixel-by-pixel basis. As seen in Fig. 3a, due to the strong decrease in the lifetime with temperature, $R_{mean}$ has a pronounced dependence on the temperature. Figure 3b shows the temperature sensitivity ($S_R$) of $R_{mean}$ evaluated from Equation (3). It is shown that $S_R$ increases from 3.5 to 11.5 %/K in the 295 - 335 K range.

$$S_R = \frac{1}{R_{mean}} \left| \left( \frac{dR_{mean}}{dT} \right) \right| \quad (3)$$

The uncertainty of the temperature measurements is evaluated from calibration data. The temporal (shot to shot) standard deviation of the measurements is 0.13 K over the 296 – 333 K temperature range, increasing to 0.76 K at 341 K. Also, the spatial standard deviation of the measurements is 0.25 – 0.5 K over the 296 – 333 K range; increasing to 3.51 K at 341 K. The increasing uncertainty with temperatures is due to decreasing signal-to-noise ratios occurring as a result of thermal quenching and the shorter phosphorescence lifetimes. As a result, the phosphorescence signal intensity in F2 (Fig. 2b) is significantly reduced at 333 K by 90% of its value at 297 K and at 341 K, by 96 %.

### 2.3. Flame front (CH*) imaging

A high-speed camera (VEO 710L, Phantom), with a 433±14 nm bandpass filter mounted on a 50 mm Nikon camera lens (f/1.2), was used to image the flame front (CH*) within the crevice. The camera imaged the light reflected by the 505 nm dichroic mirror. The camera operated at a 2 kHz frame rate and with an exposure time of 500 µs. The CH* acquisition was also synchronized with ignition timing.

The CH* imaging setup provided a ~ 190 µm/pixel resolution, giving a final spatial resolution of 380 µm/pixel after 2 × 2 pixel binning. This resolution matches that of the $T_{wall}$ measurements. The CH* and the wall thermometry images were mapped using a MATLAB function such that the pixel-pixel displacement of images from both diagnostics was equivalent to the 380 µm spatial resolution.

## 3. Results and discussion

### 3.1. $T_{wall}$ and flame distribution

Figure 4a shows instantaneous $T_{wall}$ and CH* images describing the wall temperature and flame distribution as the flame propagates in the crevice. The flame distribution is shown within the entire span of the crevice to describe its overall geometry relative to the 22 × 22 mm$^2$ phosphor area. The flame is moderately wrinkled, which varies spatially and temporally. The level of wrinkling increases with increasing pressure due to a decreased flame thickness [23]. The flame burns parallel to the crevice walls and depicts a sidewall quenching configuration.

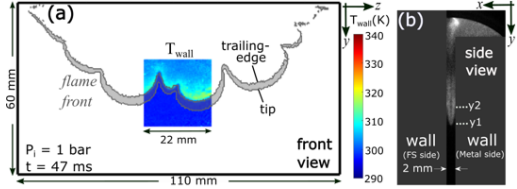

Fig. 4. (a) instantaneous $T_{wall}$ and overlaid CH* image ($P_i$ = 1 bar). (b) side-view CH* image (separate from (a)).

The wrinkled flame geometry is well captured in the $T_{wall}$ image. To understand the $T_{wall}$ distribution, it is necessary to discuss the flame geometry normal to the wall. To achieve this, additional experiments imaging CH* through the side window were conducted. Figure 4b shows an example CH* image. Note that CH* images in Fig. 4a and 4b correspond to different experiments. The flame is shown to have an "arrow-head" shape with its tip, or leading-edge, centered between each wall at location y1. The flame closest to the wall (i.e., flame's trailing-edge) is located further upstream at y2. The distance between the leading-edge and trailing-edge gives the elongated CH* signal shown in Fig. 4a. In our previous work [11,12], it was shown that $T_{wall}$ increases sharply as the flame closest to the wall (i.e., y2) sweeps over the phosphor coating. This feature is seen in Fig. 4a. Namely, $T_{wall}$ increases sharply at the trailing-edge of the CH* signal and, consequently, $T_{wall}$ conforms to flame's trailing-edge geometry.

The unique spatial features of $T_{wall}$ with respect to flame geometry is discussed further in Fig. 5. This is discussed for two single-shot images shown for different initial pressures. Figure 5a corresponds to the same high-speed imaging sequence in Fig. 4a, but shown at a different instant in time. The boundaries of the CH* signal are overlaid as white outlines onto the $T_{wall}$ images. For the images shown, the flame for $P_i$ = 1 bar exhibits large scale wrinkles with well-defined cusps and crests, while the flame for $P_i$ = 2 bar exhibits a local flame crest with mild curvature and finer scale wrinkles. These flame features are local to the TGP area at this instant in time. Different flame features can exist outside this area and the images in Fig. 5 do not imply that a flame is more wrinkled for $P_i$ = 1 bar.

$T_{wall}$ conforms remarkably to the distinct flame geometry, including large and fine scale wrinkling. $T_{wall}$ is highest directly after the CH* trailing-edge and decreases monotonically with increasing distance from the flame. $T_{wall}$ is greater for $P_i$ = 2 bar. This is because at higher pressure (pressure traces shown in Fig. 5b and e) the flame is pushed closer to the wall, which yields higher wall temperatures [3,24]. On average, $T_{wall}$ directly behind the CH* trailing-edge is 7 K greater for $P_i$ = 2 bar than 1 bar.

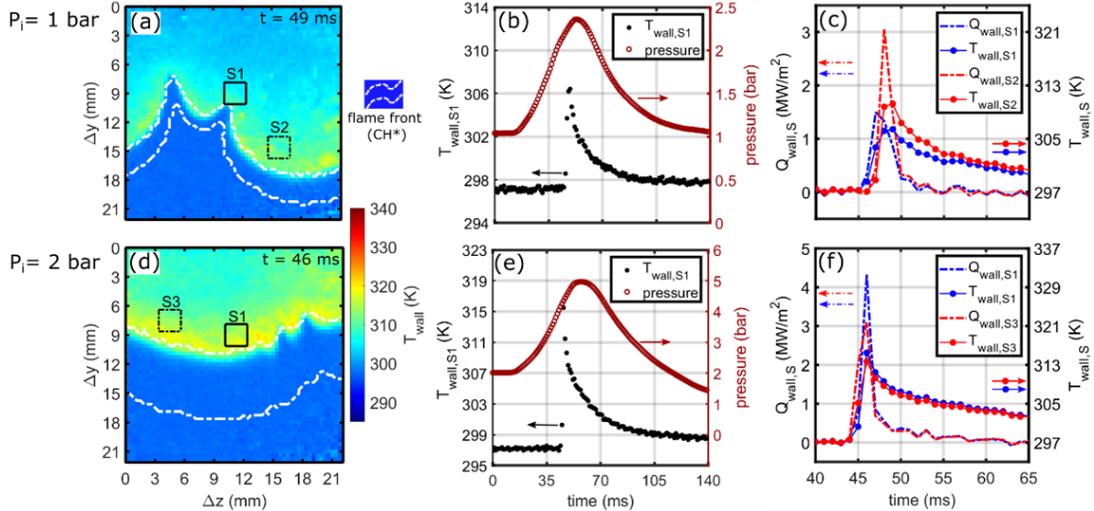

Fig. 5. $T_{wall}$ images at selected times for $P_i$ = 1 and 2 bar. Spatially-averaged $T_{wall}$ and $Q_{wall}$ are extracted from rectangles S1 – S3. Temporal evolution of $T_{wall}$ and $Q_{wall}$ are shown in the plots. Chamber operating pressure is also shown in the plots.

The spatial features of $T_{wall}$ are further evaluated in Fig. 5 plots, which reports spatially-averaged $T_{wall}$ measurements extracted from rectangles S1, S2 and S3 (each 4.6 mm$^2$ area). The plots in Fig. 5, report the time history of $T_{wall}$ at these locations. The wall heat flux, $Q_{wall}$ is also reported at these locations. $Q_{wall}$ is based on the time-dependent surface temperature of a thermally semi-infinite solid. This is treated as 1D transient heat conduction, resulting in a Duhamel integral that provides the time evolution of $Q_{wall}$ given by Equation (4) [25].

$$Q_{wall}(t) = \sqrt{\frac{\rho C_{p,w} k}{\pi}} \int_0^t \frac{[dT_{wall}(\bar{\tau})/d\bar{\tau}]}{\sqrt{t-\bar{\tau}}} d\bar{\tau} \quad (4)$$

Where $\rho$, $C_{p,w}$ and k are the density, specific heat capacity, and thermal conductivity of the wall material (stainless steel). The variable t is time and $\bar{\tau}$ is the time variable of integration. Since measurements of $T_{wall}$ are based on the temperature of the thin TGP coating, it is understood that for transient heat transfer, there may be a bias in estimating the $Q_{wall}$ from Equation (4) when considering properties of the crevice wall material (stainless steel) instead of the properties of the phosphor coating. The challenge is that the physical properties of the TGP are unknown. In this work, the TGP coating has a thickness of ~ 7 μm across which temperature gradients are considered to be negligible during transient combustion [26]. Therefore, we argue that the properties of stainless steel used in Equation (4) are sufficient to estimate $Q_{wall}$.

It is recognized that the 1 kHz repetition rate may under-resolve the maximum $Q_{wall}$. Although higher repetitions rates as shown in our previous work [12] better resolve $Q_{wall,max}$, the 1 kHz repetition rate is sufficient to compare local $Q_{wall}$ values.

Rectangle S1 has the same spatial location within both image sets. This would represent a traditional point-wise TGP measurement fixed in space [12]. S2 and S3 are chosen to compare/contrast $T_{wall}$ from S1. Each rectangle is located behind the flame location to resolve the local $T_{wall,max}$.

In Fig. 5a, S1 is located behind a flame cusp, while S2 is located near a flame crest. In this work, $T_{wall}$ behind flame cusps consistently exhibits lower temperatures than flame crests. Figure 5c shows that $T_{wall}$ is 4 K higher for S2 than S1, while $Q_{wall}$ is more than twice as large for S2. The Lewis number for the $CH_4$-air mixture is ~ 1. Thus, preferential diffusion, which would create higher (lower) flame temperatures at crests (cusps), is not expected to be significant. However, other effects such as the hydrodynamic instability [23] or greater flame distance from the wall may result in lower $T_{wall}$ for cusps than crests. This is discussed further in Sect. 3.2. In Fig. 5d, with the exception of finer-scale wrinkles (e.g. two cusps with lower $T_{wall}$), $T_{wall}$ is fairly uniform behind the mildly curved flame crest. S1 and S3 show less discrepancy in $T_{wall}$ and $Q_{wall}$ as shown in Fig. 5f.

These measurements demonstrate the unique advantage to resolve spatial variation of $T_{wall}$ and $Q_{wall}$ with respect to distinct flame features. This information is important to interpret specific findings of FWI, e.g., $Q_{wall,max}$ or flame quenching distance, which can be derived from $Q_{wall,max}$ [12,24].

## 3.2. Spatiotemporal dynamics of flame and $T_{wall}$

Figure 6 shows a temporal sequence of CH* and $T_{wall}$ to describe the spatiotemporal dynamics of the flame and wall. The images are shown every 3 ms, but a video sequence of each image can be found in the supplemental material. This image sequence is shown for $P_i$ = 2 bar.

The local flame contour in the images is noticeably wrinkled. At t = 47 ms, distinct flame cusps and crests exist and are labeled by A and B, respectively. During the image sequence, the local flame contour undergoes a sizeable cusp formation as the flame penetrates further into the crevice. This cusp formation occurs between t = 50 – 59 ms as crests B1 and B3 merge together. $T_{wall}$ is extracted along Δy = 9.4 mm (dashed-line within images) to further quantify temperature magnitudes and gradients in relation to flame features. This is shown in the bottom row of Fig. 6. The red-transparent area highlights the region with CH* signal. The edges of these regions represent the location of the CH* trailing-edge.

The $T_{wall}$ images remarkably resolve the unique spatiotemporal features of the flame. $T_{wall}$ is largest near the CH* trailing-edge where the flame is closest to the wall. $T_{wall}$ monotonically decreases with distance from the flame as the wall has more time to cool after the flame has passed. Temperature gradients as high as 26 K/mm exist for wall locations crossing the flame boundary. These large gradients exist in regions bounded by a single flame front (e.g., Δz = 14 mm, t = 47 ms), as well as within cusp regions with two surrounding flame fronts (e.g., Δz = 0-3 mm, Δz = 12 mm, t = 50 ms). A "hot spot", located at Δy = 14 mm, Δz = 13mm exists in the $T_{wall}$ images. This isolated spot (3×4 pixels, ~ 1.7 mm$^2$ in size) is due to a small bump on the TGP coating locally formed while applying the coating on the wall using the airbrush. This small bump is an imperfection, perhaps caused by an air bubble impacting the surface, which is sometimes unavoidable when applying a TGP coating. Due to its small area (1.7 mm$^2$), we could not measure this bump's thickness with the CT gauge, however, the bump is anticipated to have a thickness greater than ~ 7 ± 2 μm. After the flame passes the measurement region, this 1.7 mm$^2$ spot remains 5 – 8 K hotter than its surroundings. This feature does not exist anywhere else on the $T_{wall}$ images. Since this bump is an imperfection, the locally higher $T_{wall}$ at this location are biased and are not a unique feature of FWI. Thus, this 3×4 pixel region is ignored in the following analysis.

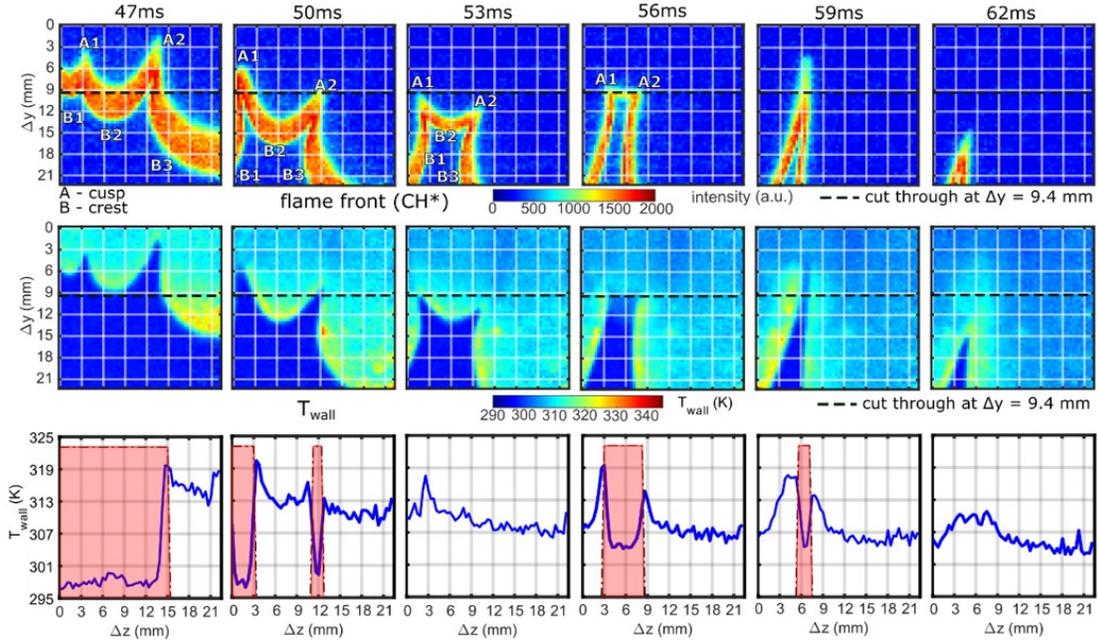

Fig. 6. Spatiotemporal sequence capturing flame / $T_{wall}$ dynamics for a highly wrinkled flame ($P_i$ = 2 bar). Flame cusps and crests are labeled by A and B, respectively. The bottom row shows $T_{wall}$ extracted along $\Delta y$ = 9.4 mm. The red area shows regions with CH* signal. The edges of these regions represent the location of the CH* trailing-edge, where the flame is closest to the wall.

$T_{wall}$ images consistently show higher temperatures associated with flame crests than for flame cusps. The higher temperatures can be seen for crests B2 and B3 where $T_{wall}$ ranges from 319 – 327 K, while $T_{wall}$ behind cusps A1 and A2 range from 305 – 312 K for this image sequence. The cooler $T_{wall}$ associated with flame cusps is most evident for the large flame cusp formed t = 56 – 59 ms. During this time, the flame cusps move upstream and the CH* signal becomes progressively thinner and weaker. The thinning of the CH* signal is associated with the flame tip and trailing-edge (see Fig. 4b) being at similar locations. This has been described in our previous work and occurs when the flame is quasi-stationary in the crevice [12]. The weaker CH* signal is particularly noticeable at the trough of the cusp at t = 56 ms. Coinciding with this weaker CH* signal is a significantly lower $T_{wall}$ of ~ 305 K. Comparison of this lower $T_{wall}$ with local and preceding values is seen in the plots of Fig. 6. For example, at t = 56 ms the cusp region is ~ 5 K colder than regions to the right where the flame passed 10 – 12 ms earlier. In addition, the cusp region at t = 56 ms is ~ 10 – 15 K colder than previous temperatures recorded at the same location from t = 50 – 53 ms, despite a flame being nearby at each of these instances in time. From t = 58 – 59 ms, the flame crests B1 and B3 merge together as the cusp closes. During this time, there is a distinct narrow channel of colder $T_{wall}$ penetrating through the trough of the cusp ($T_{wall}$ as low as 303 K), while surrounding areas exhibit significant heating from the nearby flame front with $T_{wall}$ as high as 323 K.

To understand the lower $T_{wall}$ associated with this severe flame cusp, the full 3D flame geometry, gas temperature, and flow velocity are needed. However, based on the available measurements, we offer a brief discussion that may explain these colder wall temperatures. As the flame is wrinkled, the flame will experience a hydrodynamic (or Darrieus-Landau) instability. This instability arises from the thermal expansion across the flame front [23,27]. For a wrinkled flame, this instability channels unburned fluid towards local flame cusps [27], which slows the flame propagation in cusp regions. Longer residence times will be associated with the slower flame propagation. As a result, the flame can have more time for heat transfer at the wall, which can explain the observed lower $T_{wall}$ associated with flame cusps. If severe enough, the increased heat transfer can locally quench the flame, which may explain the flame cusp at t = 56 ms where weaker CH* signal is present at the cusp region. In addition to increased heat loss, the flow velocity channelled into cusp regions imposed by the hydrodynamic instability can increase the compressive strain at the flame surface. The additional strain can lower the flame's local heat release, which may also explain the weaker CH* signal and lower $T_{wall}$ for flame cusp at t = 56 ms. The compressive strain can contribute to local flame quenching [28], which would further augment the lower CH* signals and lower $T_{wall}$. Readers are referred our previous work, which discusses CH* and $T_{wall}$ signatures associated with flame quenching in the crevice passage [12].

The $T_{wall}$ image at t = 59 ms supports the argument that a strong channelled flow of unburned gas may exist at the flame cusp. The narrow channel of colder $T_{wall}$ at t = 59 ms is suspected to be caused by a strong flow of unburned gases impinging onto the wall, which in this instance, actively cool the wall. The resulting narrow $T_{wall}$ signature is unique in that it only occurs when two flame cusps merge and leaves a distinct narrow channel of colder wall temperature at the center of the cusp. The extent to which the flow cools the wall is rather remarkable; $T_{wall}$ in this narrow channel is ~ 10-20 K colder than the surrounding enflamed regions and ~ 5 K colder than other post-burned gas regions. DNS studies employing conjugate heat transfer would be valuable to better understand the unique flame/wall dynamics associated with flame cusps in two-wall passages.

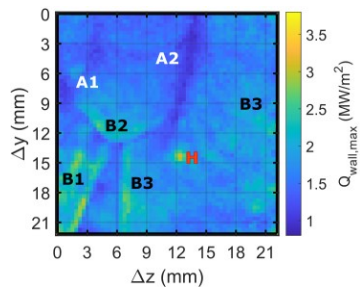

Fig. 7. $Q_{wall,max}$ distribution for image sequence in Fig. 6.

The spatiotemporal $T_{wall}$ measurements provide the unique opportunity to track the $Q_{wall}$ associated with distinct flame features. $Q_{wall}$ is calculated according to Eqn. 4, and Fig. 7 reports $Q_{wall,max}$ recorded at each pixel for the entire imaging sequence described in Fig. 6. The map of $Q_{wall,max}$ in Fig. 7 captures the distinct flame features that occurred throughout the image sequence. For example, the trajectories of the flame cusps A1 and A2 are shown by the traces of lower $Q_{wall}$ values (< 1.4 MW/m$^2$), which result from lower $T_{wall}$ at these locations. The trajectories of B2 and B3 as they travel in the measurement area are shown in between the cusp regions and on the right side of the image, respectively. These crest regions yield $Q_{wall,max}$ values nearly twice as high as cusp regions (2.0 – 2.6 MW/m$^2$). The merging of crests B1 and B3 from t = 53 – 65 ms is captured in the bottom-left region and exhibit the highest values from this imaging sequence with $Q_{wall,max}$ as high as 3.0 MW/m$^2$. The hot spot (H) is also evident in Fig. 7 with $Q_{wall,max}$ = 3.1 MW/m$^2$, but is ignored since this is argued to be a coating defect. Figure 7 also reveals discrete sections of $Q_{wall}$, which correspond to the flame location captured at the imaging timing. A higher repetition rate would provide a more continuous 2D map of $Q_{wall}$.

## 4. Conclusions

Wall temperature ($T_{wall}$) measurements and flame-wall interactions (FWI) in a narrow crevice passage have been recorded at 1 kHz using phosphor thermometry and flame front (CH*) imaging. Using the phosphor ScVO$_4$:Bi$^{3+}$, high-precision and spatially resolved $T_{wall}$ measurements uniquely captured $T_{wall}$ transients associated with FWI. The measurements reveal how the flame's spatiotemporal features influence the $T_{wall}$ distribution. This aspect is demonstrated for combustion of a CH$_4$-air ($\phi = 1.0$) charge at initial pressures of 1 and 2 bar. At higher pressure, the flame is pushed closer to the wall, leading to higher $T_{wall}$, and higher wall heat fluxes ($Q_{wall}$).

Distinct features of $T_{wall}$ and $Q_{wall}$ associated with spatiotemporal dynamics of FWI was presented. Overall, $T_{wall}$ is largest near the CH* trailing-edge where the flame is closest to the wall. $T_{wall}$ conforms remarkably to the flame geometry, including large and fine-scale wrinkling. For a wrinkled flame at these operating conditions, flame cusp regions consistently exhibit lower $T_{wall}$ and $Q_{wall}$ than flame crest regions. Findings demonstrate the utility of the imaging diagnostics to resolve spatiotemporal variation of $T_{wall}$ resulting from unique flame features, which may include transient heat transfer effects accompanying flame/flow instabilities.

## Acknowledgements

We gratefully acknowledge funding from the ERC (grant #759546) and EPSRC (EP/P020593/1, EP/V003283/1). We also thank B. Fond and C. Abram for providing the ScVO$_4$:Bi$^{3+}$ sample for this work.

## Supplementary material

Supplementary material (videos) of $T_{wall}$ and CH* imaging from Fig. 6 is provided.